\documentclass[traditabstract]{jcil}
\usepackage[utf8]{inputenc}
\usepackage[round,colon]{natbib}
\usepackage{amssymb}

\title{Conspiratorial cosmology---the case against the Universe}
\author{J\"org P. Rachen\inst{1,2}\thanks{Email: universe23@jpr-cosmic.de}
\and Ute G. Gahlings\inst{2}}
\institute{\small 
Institut f\"ur Zahlenmystik, Rautavistische Universit\"at Grafenhausen, Germany
\and Institut f\"ur angewandte Oligophrenie, Rautavistische Universit\"at
Gr\"afinnenhausen, Germany}

\date{Received April 1; Accepted April 1; Published April 1, 2013}

\begin{document}



\abstract{%
Based on the cosmological results of the Planck Mission, we show that all
parameters describing our Universe within the $\Lambda$CDM model can be
constructed from a small set of numbers known from conspiracy theory.
Our finding is confirmed by recent data from high energy particle physics. This
clearly demonstrates that our Universe is a plot initiated an unknown interest
group or lodge. We analyse possible scenarios for this conspiracy, and conclude
that the belief in the existence of our Universe is an illusion, as previously
assumed by ancient philosophers, 20th century science fiction authors and
contemporary film makers.%
}





\setcounter{page}{966}

\maketitle

\section{Introduction}

Since the dawn of culture, the believe in creation has been the foundation of
all cosmology \citep{Rigveda, Genesis, Bigbang}, although some funny ideas about
eternity \citep{Aristoteles} temporarily confused the minds of scientists
\citep[e.g.,][]{Hoyle,Bondi}. Started eventually in 1964, a joint
venture of telephone engineers and astrophysicists brought us the insight that a
signature of this creation is still around us \citep{PenziasWilson,Dicke}. It
took us another few decades to learn that this signal contains useful
information \textit{when} exactly and \textit{how} creation has happened
\citep{Cobe,WMAP_Results}. The quest to find these answers culminated in the
Planck Mission \citep{Tauber,PlanckMission}, which recently released its
results. 

A question never asked in this context, neither by modern cosmology nor by its
ancestors, is \textit{why} the Universe was created. \textit{Who} has an
interest in its existence? In all other aspects of life, preliminary answers to
such kind of questions are given by conspiracy theory, i.e., the assumption
that everything which happens is controlled by an interest group or lodge, whose
actions are generally obscure to normal creatures and manifest themselves as
mysterious ``forces'' \citep[and countless other work]{Illuminatus}. This raises
the question whether a similar logic can be applied to the Universe as a whole.

A common element of cosmology and conspiracy theory is their affection to
numbers. In modern cosmology, the entire knowledge about the Universe is cast
into a set of numbers called the \textit{cosmological parameters}, and revealing
these numbers has become the main driving force of experimental
cosmology. Conspiracy theory, in turn, assumes that at least some members of
the lodge have a favour for numbers, and have fun in continuing to communicate
them to us through shamans, mathematicians, lunatics, science fiction authors,
potheads, and other initiates.

In this paper, we explore the potential conspiratorial origin of our
Universe by showing how almost all relevant fundamental parameters can be
constructed by simple mathematical operations from a small set of 
conspiratorial numbers. As orders of magnitude and units generally play no role
in
conspiracy theory, we introduce the notation $X\cong C
\langle[{\rm unit}]\rangle$ for a conspiratorial number $C$ being
consistent with a physical quantity $X$ measured in the given ``unit'' within
the $2\sigma$ error range, after performing an arbitrary decimal shift
to $C$; for dimensionless quantities, ``[unit]'' is omitted. In case errors are
not known or for other cases of less accurate comparisons, we use the notation
$X \approxeq C \langle[{\rm unit}]\rangle$. We refer to these relations as
\textit{conspiratorial correspondence} in the narrow and wide sense,
respectively. We adopt the usual convention to refer to the Conspirators by
using upper-case pronouns.

\begin{table*}[t]
\begin{tabular}{lcc|cc}
\hline
\multicolumn{2}{c}{base parameter} & $\cong$ & $\sigma/\Pi$ & $N_c$
\\\hline
\makebox[0.58\textwidth]{Physical baryon density\dotfill} & 
\makebox[0.15\textwidth]{$\omega_b \equiv \Omega_b h^2$\dotfill}
& $23{c}$ & $6.9\,10^{-3}$ & $11$ \\ 
\makebox[0.58\textwidth]{Scaled physical matter density\dotfill} & 
\makebox[0.15\textwidth]{$\Omega_m h^3$\dotfill}
& ${c}$ & $8.9\,10^{-4}$ & $11$ \\ 
\makebox[0.58\textwidth]{Redefined acoustic scale measure\dotfill} & 
\makebox[0.15\textwidth]{$\theta_*'$\dotfill}
& $42$ & $3.4\,10^{-3}$ & $11$ \\ 
\makebox[0.58\textwidth]{Thomson scattering optical depth due to
reionization\dotfill} & 
\makebox[0.15\textwidth]{$\tau$\dotfill}
& ${c}$ & $4.8\,10^{-2}$ & $11$ \\ 
\makebox[0.58\textwidth]{Scalar spectrum power law index \dotfill} & 
\makebox[0.15\textwidth]{$n_s$\dotfill}
& ${c}$ & $4.7\,10^{-2}$ & $1$ \\ 
\makebox[0.58\textwidth]{Log power of the primordial curvature
perturbations\dotfill} & 
\makebox[0.15\textwidth]{$\ln(10^{10}A_s)$\dotfill}
& $\pi$ & $5.5\,10^{-2}$ & $2$ \\ \hline
\end{tabular}
\caption{\label{planckbase} Conspiratorial correspondence of modified Planck
base
parameters (see text). The parameters $\sigma/\Pi$ and $N_c$ are needed to
determine the statistical significance of this finding (see Eq.~\ref{pvalue}).}
\end{table*}

\section{Conspiratorial numbers}

\subsection{23}

The smallest prime number which is the sum of three consecutive prime numbers
is $23 = 5 + 7 + 11$. It is also the only integer number bracketed by $\pi^e$
and $e^\pi$ (Scott, priv.~comm.).\footnote{While $\pi$ is a conspiratorial
number beyond doubt (see Sect \ref{pi}), the role of the Euler number $e$ in
conspiracy theory is still under debate and subject to intense research. It is
therefore not considered in this paper.} It is the foremost number of conspiracy
theory. According to tradition, the origin of the 23-enigma is attributed to the
US author and pop icon William S. Burroughs \citep[see, e.g.,][]{Wilson23}. It
has been spread through standard work of conspiracy theory
\citep{Illuminatus,CosmicTrigger}, and nowadays fills countless blogs and
web-pages of paranoid conspiracy fans. It is therefore obvious that no
number-based conspiracy theory can be constructed without this number. 

\subsection{42} 

Ever since its proposal as ``The Answer'' by \citet{Adams}, this number
has entered a fixed place in the thinking of a whole generation of scientists.
It connects scientific methodology, i.e., analysing numerical results
without knowing which question has been asked; creation, as it turns out
that our world was \textit{built} to find that question; and conspiracy, as
there was apparently some disagreement whether the creation of the Universe was
good move \citep[Vol~2, Chap~1]{Adams}. Moreover, as independently
noted by Knoche and Scott (both priv.~comm.), $42$ written to base $2$ means $3$
times on-off (i.e., $101010$), which reveals again $23$.

\subsection{$\pi$}\label{pi}

Traditionally associated to circles \citep{Archimedes}, the conspiratorial
nature
of $\pi$ becomes obvious only when we consider that its distinguished
geometrical meaning occurs only in flat space \citep{Euclid}, which has been
turned out to be one gross misrepresentation: First we had to learn that the
surface of Earth is not flat \citep{Eratosthenes}, then that space itself is
curved in almost every place of interest \citep{Einstein}. Only very recently,
it turned out that the abstract concept of empty space in our Universe is indeed
Euclidian to high precision \citep{PlanckMission}, which lets us conclude that
some obviously inaugurated ancient Greeks knew something which the rest of
us needed at least $23$ centuries to figure out. 


\section{Conspiratorial cosmology}

\subsection{Conspiratorial values for physical parameters} 

Conspirators are malicious, but They are not subtle. We can therefore assume
that the construction of the cosmological parameters out of the conspiratorial
numbers has to follow simple mathematical operations, such as multiplying
them with each other. Following this principle, we construct the conspiratorial
values $23\pi$, $42\pi$ besides the fundamental conspiratorial numbers $23$,
$42$ and $\pi$, and introduce the superconspiratorial constant 
${c} = 23\times 42 = 966$.\footnote{To avoid confusion, we note that in the
physics literature $c$ is occasionally used to denote the velocity of light.}
Without considering high precision CMB data, these parameters
seemed to suffice to represent the most fundamental parameters of the
$\Lambda$CDM cosmology, as the dark energy density $\Omega_\Lambda \cong 23\pi$,
baryon density $\Omega_b\cong 42$, dark matter density $\Omega_c \cong 23$, and
Hubble constant $H_0 \cong 23\pi$ [km/s/Mpc] \citep{SN1}. Moreover, CMB data
obtained by WMAP already showed that the primordial spectral index $n_s \cong c$
\citep{WMAP_Parameters}.

Following the principle of complexification described by \citet[Vol~2,
Intro]{Adams}, we have to expect that this cannot hold for measurements at
higher precision, and the Planck results will require refinements
in order to reveal their conspiratorial nature. Here we note that the
natural conspiratorial symmetry ${c} \cong 1$ occurring for sufficiently large
errors may break by refined measurements, so we expect $c$ to be the 
correction factor needed to bring these measurements in line. Thus we allow all
conspiratorial values to be multiplied with ${c}$, except ${c}$ itself as
squared superconspiracy is imbecilely unstable (Rachen and Gahlings, in
preparation). This defines $11$ conspiratorial values which we compare with the
Planck results. 

\begin{table*}[t]
\begin{tabular}{lcccc}
\hline
\multicolumn{2}{c}{derived parameter} & $\cong$ & $\approxeq$ & 
\\\hline
\makebox[0.58\textwidth]{Dark energy density divided by critical
density today\dotfill} & 
\makebox[0.15\textwidth]{$\Omega_\Lambda$\dotfill}
& $23\pi{c}$ & $23\pi$ \\ 
\makebox[0.58\textwidth]{Matter density today decided by critical
density\dotfill} & 
\makebox[0.15\textwidth]{$\Omega_m$\dotfill}
& $\pi$ & $\pi{c}$ \\ 
\makebox[0.58\textwidth]{Current expansion rate in km/s/Mpc\dotfill} & 
\makebox[0.15\textwidth]{$H_0$\dotfill}
& $23\pi{c}$ & $23\pi$ \\ 
\makebox[0.58\textwidth]{Redshift at which the Universe is half
reionized\dotfill} & 
\makebox[0.15\textwidth]{$z_{\rm re}$\dotfill}
& --- & ${c}$,$42\pi{c}$ \\ 
\makebox[0.58\textwidth]{BAO distance ratio at $z=0.57$\dotfill} & 
\makebox[0.15\textwidth]{$r_{\rm drag}/D_V(0.57)$\dotfill}
& $23\pi$ & $23\pi{c}$ \\ 
 \hline
\end{tabular}
\caption{\label{planckderived} Conspiratorial correspondence of Planck derived 
parameters (see text). For each parameters, the best and second best match
among the possible conspiratorial values is shown. For $z_{\rm re}$ there is no
correspondence in the narrow sense, but two in the wider sense.}
\end{table*}

\subsection{Comparison with Planck results}

Following \citet{PlanckParameters} we distinguish between \textit{base
parameters} directly determined from CMB maps, and \textit{derived parameters}
within the $\Lambda$CDM model. Close inspection reveals, however, that some of
the base parameters have been badly chosen by \citet{PlanckParameters}, so we
decide to replace $\omega_c = \Omega_c h^2$ by the much better constrained
parameter $\Omega_m h^3$, and redefine the acoustic scale as $\theta_*' \equiv
100\theta_* -1$ (for those who find this definition dubious we recall that They
\textit{are} malicious). 

The result is shown in Table \ref{planckbase}. We see that all base
parameters chosen this way show conspiratorial correspondence in the narrow
sense. To asses the statistical significance of this finding, we define for each
parameter the quantity $\sigma/\Pi$, i.e., the ratio of the determined
1-$\sigma$ error to its prior range given in \citet{PlanckParameters}, and
note that the chance probability of a conspiratorial correspondence in the
narrow sense is given for each parameter $i$ by
\begin{equation}\label{pvalue}
 p_i < 4 N_c \frac{\sigma_i}{\Pi_i}
\end{equation}
where $N_c$ is the number of conspiratorial values in the prior range, and the
inequality expresses that $p_i$ may be overestimated due to non-considered
overlaps of the error ranges around the conspiratorial values. For the total
chance probability that the match shown in Table \ref{planckbase} is purely
coincidental (i.e., non-conspiratorial) is then $p = \prod_i p_i <
1.5\,10^{-4}$,
which clearly exceeds the conspiratorial confidence threshold of 23 decisigma.

We note that not all parameters contribute to this significance. For the optical
depth $\tau$ the error bars are so large that essentially any possible value
could have been interpreted as a conspiratorial match. For the more tightly
constrained combination with the matter density perturbation power, however, we
find $\sigma_8\,e^{-\tau} \approxeq 23\pi$. Efstathiou (priv.~comm.) pointed
out that also $\sigma_8^2 \cong \pi/42$, but the significance of such more
subtle correspondence needs further investigation. Eventually,
Table~\ref{planckderived} shows that a significant fraction of the derived
$\Lambda$CDM parameters show conspiratorial correspondence, among them those
which capture the highest public interest, like $\Omega_\Lambda$ and $H_0$. It
should be obvious that this is not above board. 

\section{Discussion}

\subsection{Other fundamental parameters}

Besides CMB physics, also high energy particle physics exhibits fundamental
aspects of the Universe. Of course, we cannot expect that the myriads of
particle masses or quantum numbers are represented by conspiratorial
numbers---for sure They are not na\"ive. Rather, it seems that They hide their
message only in the most fundamental principles, such as spontaneous symmetry
breaking. Indeed, the Cabibbo parameter $\lambda = \sin\,\theta_c \cong 23 {c}$
to high precision \citep{PDG}. The clearest hint of conspiracy in fundamental
physics is given by the recent discovery of the Higgs particle at a mass $m_H
\approxeq 42\pi{c}$ [GeV] \citep{HiggsCERN}. Clearly, this aspect deserves
deeper investigation with a watchful eye.\footnote{We note in this context that
the Higgs field was proposed in 1964 \citep{Higgs1,Higgs2}.}

\subsection{Previous work}

Pioneering work on conspiratorial cosmology has been done by \citet{Scott2}, who
noticed already that cosmological parameters exhibit some 
numerical correspondences. We confirm this finding by showing that
conspiracy in cosmological parameters delivered by Planck is not
homogeneous, but clusters around very few specific conspiratorial values (see
Tables \ref{planckbase} and \ref{planckderived}).

The work of \citeauthor{Scott2} was preceded by a seminal paper by \citet{Hsu},
who proposed that the CMB would be the ideal medium for the Creator to
communicate a message to the inhabitants of His creation. \citet{Scott1}
showed that Her message would appear different to different observers,
anticipating advertising methods currently explored by an incredibly large
(${\sim}\,10^{100}$) internet company. 

It is mandatory in this context to mention also the important progress in
modern creation theories \citep[e.g.,][and their followers]{Johnson,Dembski}.
Using a methodology very similar to ours, they try to deliver evidence that the
Universe was made by ``Intelligent Design''. Our results confirm the latter of
these two assumptions.

\subsection{Plot scenarios}

There are two main scenarios for a conspiratorial creation of our Universe. 

The first scenario is that our universe was physically created by Them,
potentially in a collider experiment. As the time scales involved are quite
large, it would be questionable whether They still follow the progress of Their
experiment. It is conceivable, however, that their life time scales are
significantly different from ours, and such terrible miscalculation of scale
has readily been reported \citep[Vol~1, Chap~31]{Adams}. In this case, they
may indeed still be there and potentially establish contact with us. The science
fiction literature has conceived several ways to do this, e.g., by sending
construction plans for transport machines on the frequency  
$HI\times \pi$ to the VLA \citep{contact}, or by hiding black slabs on the
moon, which of course must have the dimensions $\pi$:23:42 (and not 1:4:9 as
wrongly predicted by \citealt{2001}).\footnote{We note that the
idea to this famous science fiction story was born at a meeting between Arthur
C. Clarke and Stanley Kubrick which took place on April 23, 1964
\citep{Clarke2}.} 

A second, in our view more likely scenario is that there is no ``Universe'' in
the regular sense, which we could observe. The general idea that reality might
be an illusion is not exactly new \citep[e.g.,][]{Buddha, Plato}, but it took
until 1964 that this was cast into a language understandable to the
technical-scientific society through a novel in which the protagonist discovers
that our world is just a computer simulation \citep{simulacron3}. Probably this
novel was an pre-release violating Their publication policy, as They managed to
prevent a wide spread of the idea by keeping the novel and an early TV
adaptation\footnote{Welt am Draht, German TV screenplay directed by Rainer
Werner Fassbinder, 1973.} largely unsuccessful.\footnote{Moreover, both the
author of the novel and the director of the TV screenplay were expelled from the
simulation within a dozen years after their respective pre-releases.} An policy
change, however, seemed to have the intent to prepare us for the discovery now
made. First, a group of French philosophers \citep[et~d'autres]{Derrida,
Focault, Lyotard, Baudrillard} made denial of reality the mainstream of
intellectual thinking in vitually all areas \citep[e.g,][]{Butler1, Butler2,
May}. Second, starting at end of the last millennium the US film industry
hammered the idea into the minds of the general public by a series of action
movies, the most famous one recalling a well-known rectangular scheme of
numbers. Eventually, the idea entered ostensibly non-fictional science through a
search for signatures of the lattice spacing used in the simulation of our
Universe in the spectrum of ultra-high energy cosmic rays \citep{Beane}. 

An appealing aspect of the latter scenario is that it is much easier, even
likely that They keep permanent contact to us, either through ``Contact Units''
or by direct projection of themselves into our world \citep{simulacron3}. This
would explain why and how the conspiratorial numbers have been repeatedly
brought to our attention, and we would expect that most of the people named in
the reference list of this paper in fact belong to Them. 

\subsection{The end of the World}

If our Universe is an experiment, it is legitimate to ask when it is expected
to be finished, in particular in view to future plans or wishes we might have
(like ``I always wanted to see Norway''). In the first scenario, cosmological
models seem to give us little constraints on this, although we cannot exclude
that They are able to change ad-hoc some of the physical parameters governing
our Universe, which could have dramatical consequences
\citep{phasetransitions}. 

In the second scenario, the situation is much more worrying as terminating our
``Universe'' would not take Them more than pulling a plug. Apart from the fact
that Doomsday may, in this case, be either totally unspectacular or completely
weird, we may also consider the possibility that the dates of this event are
somehow encoded in our consciousness. According to recent rumours we were
informed about that date by a calendar \citep{mayacalendar} attributed
to a fictional culture They have implemented in our collective memory under the
name ``Maya''.\footnote{A well chosen name, as the Sanskrit word
\textit{m\=ay\=a} means \textit{illusion} and is used in this sense in the
Buddhist literature.} As the predicted date, December 21, 2012, has passed
without noticed effects, and error is inconceivable, we have to conclude that
our Universe \textit{did end} at this date, but meanwhile They received a
funding extension and the simulation is restarted with all experiences about the
temporary shut-off erased from our memory. This incontrovertible finding
confirms us in our belief that the second scenario is the right one, and
therefore it seems to be in our interest to continue providing Them with useful
results. 

\section{Conclusions}

Following the logic of conspiracy theory, we provided compelling evidence
that our Universe was created by conspiracy. The belief in its---thus
our---existence is herewith proven to be an illusion. Some open questions
remain, for example, (a) why the distribution of applied conspiratorial values
is inhomogeneous, (b) whether and how our Universe will change after this
discovery, and (c) what did \textit{really} happen in the year 1964? More data
and improved models are expected to provide answers in about one year from now.




\begin{acknowledgements}
We thank the Planck Collaboration for their sympathy. We acknowledge spiritual
support by the omnipotent Dada and by the Rautavistik movement
(http://www.rautavistik.de/).
\end{acknowledgements}


\raggedright

\bibliographystyle{jcil}

\bibliography{consp23.bib}



\end{document}